\documentclass[11pt]{article}
\pdfoutput=1 

\usepackage{jcappub} 
\usepackage[list-units=single]{siunitx}
\usepackage{physics}
\usepackage{xcolor}
\usepackage[T1]{fontenc} 
\usepackage[shortcuts]{extdash}
\usepackage[normalem]{ulem}
\usepackage[nameinlink]{cleveref}
\usepackage{changes}

\providecommand{\sorthelp}[1]{}

\begin{document}

\title{Constraints on cosmic birefringence using \textit{E}\=/mode polarisation}

\author[1]{Arefe Abghari,\note{Corresponding author.}}
\author{Raelyn M. Sullivan,}
\author{Lukas T. Hergt,}
\author{and Douglas Scott}
\affiliation{University of British Columbia, Vancouver, BC V6T1Z1, Canada}

\emailAdd{arefeabghari@phas.ubc.ca}
\emailAdd{rsullivan@phas.ubc.ca}
\emailAdd{lthergt@phas.ubc.ca}
\emailAdd{dscott@phas.ubc.ca}

\abstract
{A birefringent universe could show itself through a rotation of the plane of polarisation of the cosmic microwave background photons.  This is usually investigated using polarisation $B$ modes, which is degenerate with miscalibration of the orientation of the polarimeters. Here we point out an independent method for extracting the birefringence angle using only temperature and $E$\=/mode signals. We forecast that, with an ideal cosmic-variance-limited experiment, we could constrain a birefringence angle of \ang{0.3} with $3\,\sigma$ statistical significance, which is close to the current constraints using $B$ modes.  
We explore how this method is affected by the systematic errors introduced by the polarisation efficiency. 
In the future, this could provide an additional way of checking any claimed $B$\=/mode derived birefringence signature.}
\notoc

\maketitle
\flushbottom

\section{Introduction}
\label{sec:intro}

The standard picture of cosmology, known as the $\Lambda$CDM model, explains most cosmological observations remarkably well \cite{OlivePeacock,LahavLiddle}, with the precision driven by measurements of the cosmic microwave background (CMB)~\cite{ScottSmoot}. CMB power spectrum estimates made by satellites such as \textit{WMAP}~\cite{Hinshaw2012} or \textit{Planck}~\cite{planck2016-l06} and ground-based experiments such as \textsc{ACT}~\cite{Aiola2020TheParameters} or \textsc{SPT}~\cite{Bianchini2019ConstraintsSpectrum} have reached nearly cosmic-variance-limited sensitivity for temperature anisotropies at arcminute scales and above. However, measurements of CMB polarisation, through the geometrical components known as $E$  and $B$ modes (see e.g., \cite{KKS1997,ZS1997,HW1997, 1997Kam, 1997Sal}), currently only provide moderate additional constraints on cosmological parameters~\cite{planck2016-l01}.

Forthcoming CMB surveys are designed to make more sensitive polarisation measurements, enabling us to improve current constraints on cosmology, including investigation of extensions to the standard model~\cite{SimonsObservatory:2018koc,LiteBIRD:2022cnt,Abazajian:2019eic,Moncelsi:2020ppj}. One of these effects, which is the subject of this paper, is cosmic birefringence. It occurs when there is a parity-violating field in the Universe that couples to electromagnetism. \cite{Carroll1990LimitsElectrodynamics,Lue1998CosmologicalInteractions}. Such a field rotates the plane of polarisation of the CMB photons as a function of the distance travelled. This rotation leads to a mixing of $E$  and $B$ modes and rotates the six CMB power spectra in the following way \cite{planck2014-a23}: 
\begin{equation}
\begin{aligned}
\widetilde C_{\ell}^{ T T} &=C_{\ell}^{T T} ; \\
\widetilde C_{\ell}^{E E} &=C_{\ell}^{E E} \cos^{2}(2 \alpha)+C_{\ell}^{B B} \sin^{2}(2 \alpha)  ; \\ 
\widetilde C_{\ell}^{ B B} &=C_{\ell}^{E E} \sin^{2}(2 \alpha)+C_{\ell}^{B B} \cos^{2}(2 \alpha) ; \\ 
\widetilde C_{\ell}^{ T E} &=C_{\ell}^{T E} \cos(2 \alpha)  ; \\ 
\widetilde C_{\ell}^{ T B} &=C_{\ell}^{T E} \sin(2 \alpha); \\ 
\widetilde C_{\ell}^{ E B} &=\frac{1}{2}\left(C_{\ell}^{E E}-C_{\ell}^{B B}\right) \sin(4 \alpha). 
\label{eq:biref}
\end{aligned}
\end{equation}
Here $\widetilde C_\ell$ refers to the observed power spectra in the presence of birefringence, while $C_\ell$ are the spectra that are observed in the absence of birefringence (or any other parity-violating effect). 
Here, we have ignored terms containing $C_{\ell}^{ E B}$ or $C_{\ell}^{ T B}$, since theory predicts them to be zero due to parity violation~\cite{Feng2006SearchingBOOMERANG,Lue1998CosmologicalInteractions,Gluscevic2010}. 
Although in principle one could imagine the birefringence being different in different directions~\cite{Kamionkowski2010,Gluscevic2012,Contreras2017,2020Gruppuso}, we only focus on isotropic birefringence in this study, i.e., we assume that the rotation angle~$\alpha$ is the same in all directions. 

As shown in \cref{eq:biref}, this effect not only changes the $EE$ and $BB$ power spectra, but also generates a non-zero cross-correlation between these modes. 
Hence, one can see that by measuring $EE$, $BB$, and $EB$ it is possible to determine the birefringence angle~$\alpha$.  Indeed, most discussions of birefringence have focused on this approach~\cite{Minami2020NewData,2009,2011,planck2014-a03,2020At1,2020At2,diegopalazuelos2022cosmic,Komatsu:2022nvu,Eskilt:2022wav}.  However, a miscalibration of the orientation of the polarimeters would have the same effect on these spectra. Thus, for a cosmological background, it is not possible to distinguish birefringence from the systematic uncertainty of a miscalibration of the orientation angle. 
One way to break this degeneracy is to use the Galactic foreground emission (see e.g., \cite{Minami2019SimultaneousExperiments,Minami:2020fin}). Since photons from foreground emission have not propagated over cosmological distances, they are not strongly affected by cosmic birefringence. On the other hand, miscalibration of detector orientation alters the foreground and the CMB spectra in the same way. Hence one could use data at different frequencies to separate birefringence from foreground and calibration effects.  Nevertheless, this approach still requires a detailed understanding of the foregrounds. 

Here we study a similar but independent method for determining the birefringence angle~$\alpha$ using only CMB temperature maps, $E$ modes, and their cross-correlation. We also examine the degeneracy between this method and another systematic effect, namely the polarisation efficiency, which characterizes the sensitivity of an antenna for measuring the polarised power of an incident field. There has been at least one attempt at constraining the birefringence angle using only the temperature and cross-correlations, but without consideration of the polarisation efficiency \cite{Gruppuso2016}.
The $E$ modes are also affected by angle miscalibration, and so in general one would have to account for all forms of systematic uncertainty simultaneously. For the sake of simplicity, throughout this paper we consider the case where any miscalibration of the orientation of the polarimeters has already been fixed using $B$ modes. 
As will be shown, polarisation efficiencies affect the $T$, $E$, and $B$ modes 
differently than the angle miscalibration.

Generally, the power received by a detector in terms of Stokes parameters is~\cite{leach2008} 
\begin{equation}
P=G(I+\rho(Q \cos 2 \delta+U \sin 2 \delta)+\xi V).
\label{eq:detector}
\end{equation}
Here $I$, $Q$, $U$, and $V$ are the Stokes parameters, $G$ is a gain factor, $\rho$ is the linear polarization efficiency, $\delta$ is the detector polarisation orientation in the coordinates used to define ($Q$, $U$), and $\xi$ represents the response to circular polarisation.

The linear polarization Stokes parameters~$Q$ and~$U$ relate to multipole coefficients by~\cite{ZS1997} 
\begin{align}
a_{\pm2, l m}=\int d \Omega \ {}_{\pm2} Y_{l m}^{*}(\hat{\boldsymbol{n}})(Q\pm i U)(\hat{\boldsymbol{n}})
\end{align}
and the power spectra for temperature and polarisation in terms of multipole coefficients are~\cite{KKS1997}
\begin{equation}
\left\langle a_{(\ell m)}^{X *} a_{\left(\ell^{\prime} m^{\prime}\right)}^{X'}\right\rangle=C_{l}^{XX'} \delta_{l l^{\prime}} \delta_{m m^{\prime}},
\end{equation}
where $X$ stands for $T$, $E$, or~$B$.
Thus, we can say that the measured $C_\ell^{TE}$ in the presence of this systematic effect is proportional to the linear polarisation efficiency~$\rho$, while $C_\ell^{EE}$ is proportional to~$\rho^2$. This effect is similar to birefringence, since in the birefringence case we would measure $C_\ell^{TE}$ reduced by $\cos(2\alpha)$, whereas $C_\ell^{EE}$ would be smaller by $\cos^2(2\alpha)$ (see~\cref{eq:biref}). 
Note the different behaviour between the two mentioned systematic effects. While the miscalibration angle generates $B$ modes from $E$ modes, leading to a non-zero $EB$ and $TB$ signals, the polarisation efficiency, simply reduces the observed amplitude of the $E$ modes and $B$ modes, but does not generate any $EB$ or $TB$ signals.

All other things being equal, the method we describe here is not as sensitive as using $B$ modes. However, given the difficulty in determining a possible birefringence angle, it is important to use all available information, and in particular to provide independent tests, which is what the $E$\=/mode approach gives us.

We have structured this paper as follows. In \cref{sec:Fish} we first investigate, using a Fisher matrix formalism, how accurately an experiment could constrain the birefringence angle through measuring only the $TT$, $TE$, and $EE$ power spectra. This analysis gives us the smallest achievable uncertainty on the birefringence angle in an ideal (cosmic-variance-limited) experiment.  In \cref{sec:MCMC} we implement a Markov chain Monte Carlo (MCMC) method to obtain the $\SI{1}{\sigma}$ uncertainty on the birefringence angle~$\alpha$ using the likelihood from real \textit{Planck} data.  Finally, we discuss how this method can be used as an additional test of birefringence in the future.

\section{Fisher forecasts}
\label{sec:Fish}

The Fisher information matrix provides a powerful formalism for quantifying the ability of an experiment to constrain a set of parameters. 
Under the assumptions of Gaussian perturbations the Fisher matrix for CMB temperature and polarisation anisotropies is given by \cite{Eisenstein1998CosmicSurveys}
\begin{equation}
    \label{eq:fisher}
    F_{i j}=\sum_\ell \sum_{X,Y}\pdv{C_\ell^{X}}{\theta_{i}}\left(\vb{M}_\ell\right)_{XY}^{-1} \pdv{C_\ell^Y}{\theta_{j}},
\end{equation}
where $C_\ell^X$ and $C_\ell^Y$ are the power in the $\ell$th multipole for $X,Y= TT, EE, TE$, $\vb{M}_\ell$ is the covariance matrix, and $\theta_i$ are cosmological parameters. We focus on the six standard cosmological parameters, \{$\Omega_\mathrm{b} h^2$, $\Omega_\mathrm{c} h^2$, $\theta_\ast$, $n_\mathrm{s}$, $A_\mathrm{s}$, $\tau$\} plus the birefringence angle~$\alpha$. The Cramer-Rao inequality puts a lower bound on the variance of an unbiased parameter, which is given by
\begin{equation}
    \sigma_{\theta_i} = \sqrt{F_{ii}^{-1}}.
\end{equation}

Using \cref{eq:biref}, and assuming that the $BB$ power is zero, we can analytically find the derivatives of the $TT$, $EE$, and $TE$ power spectra with respect to the angle~$\alpha$. For small angles, we can expand this to first order in $\alpha$ to yield
\begin{equation}
    \begin{aligned}
    \pdv{\widetilde C_\ell^{TT}}{\alpha} &= 0, \\
    \pdv{\widetilde C_\ell^{EE}}{\alpha} &= -4 C_\ell^{EE} \cos(2\alpha) \sin(2\alpha) \approx -8 C_\ell^{EE} \alpha, \\
    \pdv{\widetilde C_\ell^{TE}}{\alpha} &= -2 C_\ell^{TE} \sin(2\alpha) \approx -4 C_\ell^{TE} \alpha.
    \label{eq:deriv}
    \end{aligned}
\end{equation}

\begin{figure}[tb]
    \centering
    \centering
    \centering
    \includegraphics[scale=0.6]{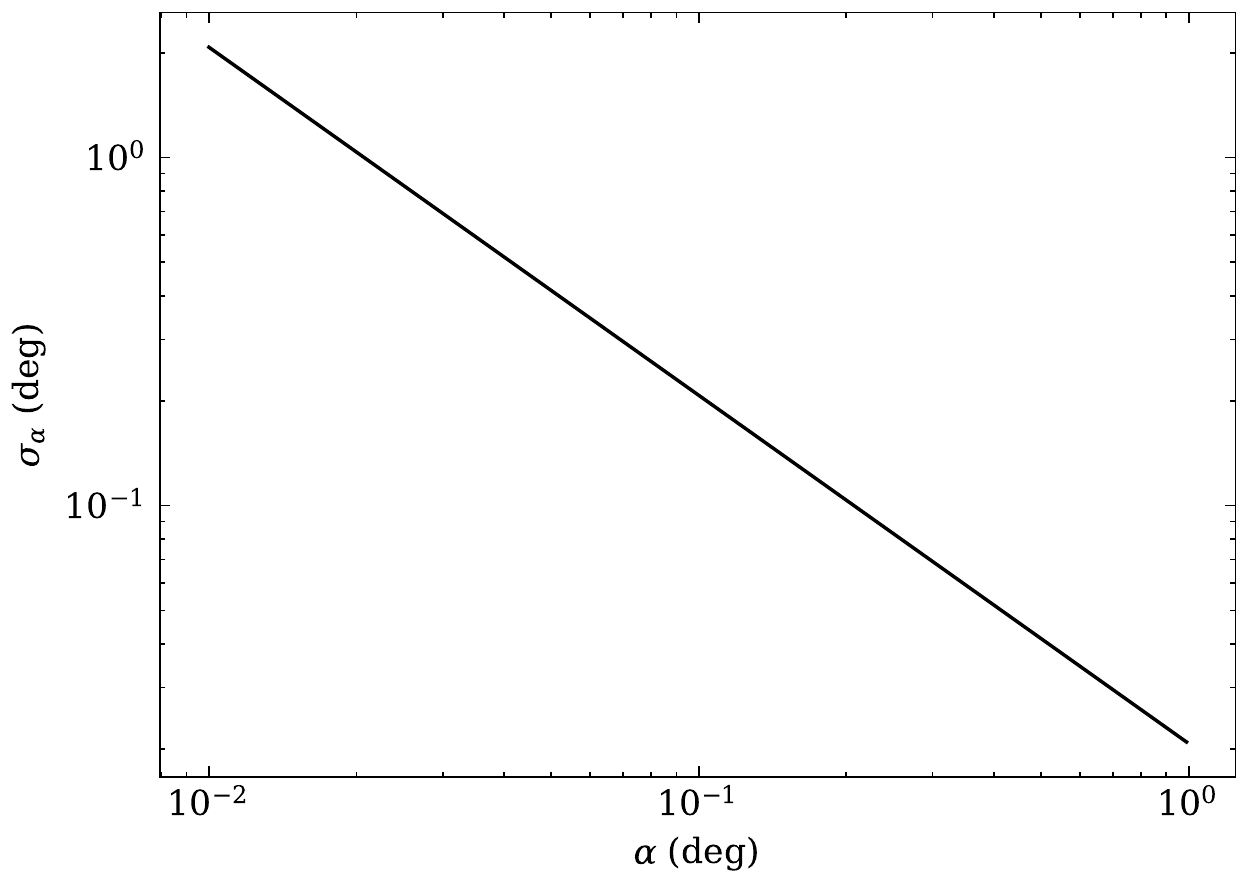}
    \caption{Uncertainty (1\,$\sigma$) on the birefringence angle~$\alpha$ obtained from a Fisher matrix calculation based on characteristics of an ideal experiment described in the text. The uncertainty derived using temperature and $E$\=/mode signals is proportional to $\alpha^{-1}$. }
    \label{fig:fisher1}
\end{figure}

The fiducial cosmological parameters used for the Fisher matrix calculation are the best-fit values from the \textit{Planck}-2018 TT+TE+EE likelihood ($\Omega_{\rm b} h^2 = 0.022377$, $\Omega_{\rm c} h^2 = 0.12010$, $100\theta_\ast = 1.04092$, $n_{\rm s} = 0.9659$, $\ln(10^{10}A_{\rm s}) = 3.0447$, $\tau = 0.0543 $) \cite{planck2016-l06}.
\Cref{eq:deriv} shows that the Fisher matrix elements related to $\alpha$ are themselves proportional to $\alpha$ and therefore its uncertainty is proportional to $\alpha^{-1}$ (see \cref{fig:fisher1}). Hence, we cannot place a bound on the statistical uncertainty $\sigma_\alpha$ if we use a fiducial model with $\alpha=0$. To treat this we choose a non-zero value for the angle~$\alpha$ in the Fisher analysis.
We compute the uncertainty derived from the Fisher matrix for different angles~$\alpha$ and consider values that would be distinguishable from the case of zero signal.

We run our Fisher analysis code both using the characteristics of the \textit{Planck} satellite and for a much more ambitious experiment with negligible noise limited by cosmic variance down to small angular scales.
For \textit{Planck}, we allow accuracy up to the multipole moment of $\ell_{\rm max} = 2500$. We use a beam size of 5.5 arcminutes and noise levels for temperature and polarisation of 11.7\,$\mu$K and 24.3\,$\mu$K, respectively. These values along with the covariance matrix elements are explicitly taken for the single \textit{Planck} channel at \SI{217}{\giga\hertz} from Ref.~\cite{Eisenstein1998CosmicSurveys}.
For an ideal future experiment, we assume that temperature anisotropies can be measured out to $\ell_{\rm max} = 3000$ and polarization out to
$\ell_{\rm max} = 6000$ (since the limiting foregrounds for temperature are expected to be absent for polarization).  We ignore beam and noise in this case,
assuming that the experiment is limited purely by cosmic variance.

\begin{figure}[tb]
\centering
\includegraphics[scale=0.7]{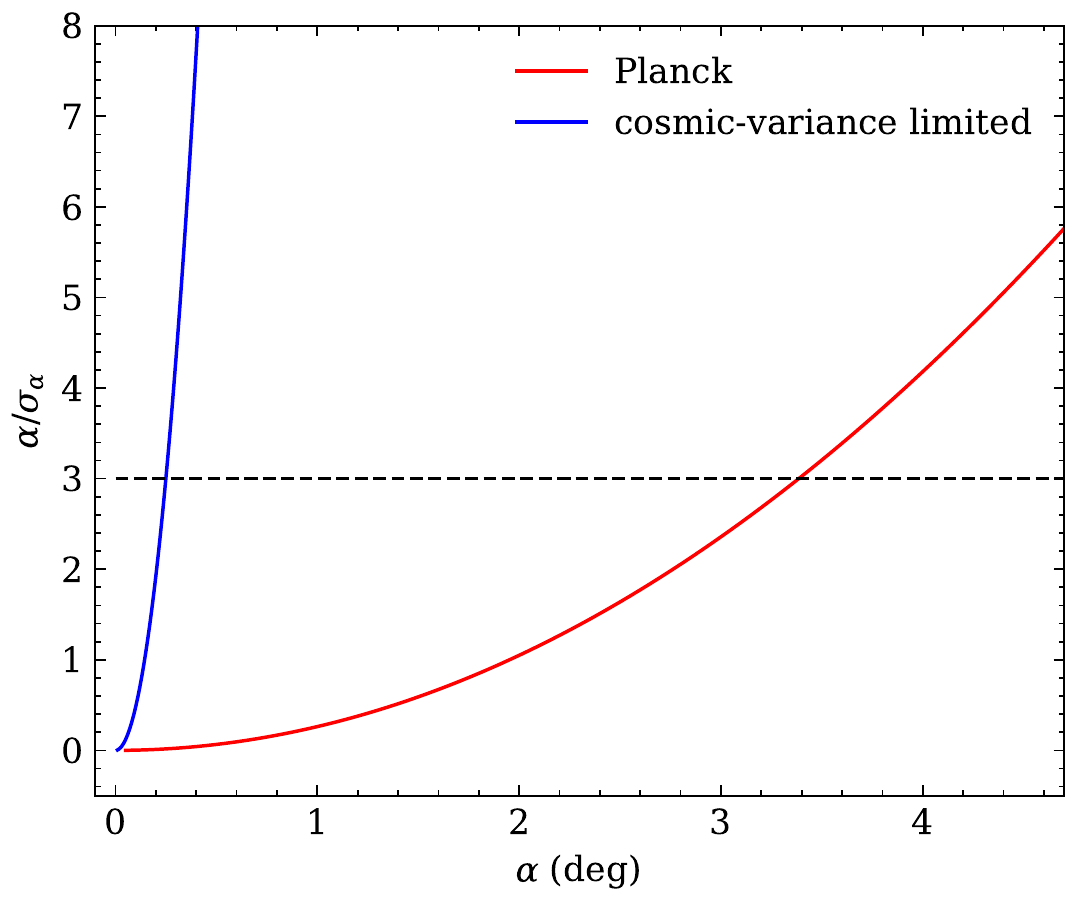}
\caption{Signal-to-noise ratio for different $\alpha$ values. The dashed line marks a signal-to-noise ratio of 3. With the \textit{Planck} experiment, for $\alpha$ values greater than \ang{3.4} the birefringence angle could be constrained with an uncertainty greater than $3\,\sigma$. For a cosmic-variance-limited experiment (neglecting beam and noise effects) going out to much higher multipoles one could detect $\alpha>\ang{0.25}$ with $3\,\sigma$ significance.}
\label{fig:fisher2}
\end{figure}

As shown in \cref{fig:fisher2}, for \textit{Planck}, we find an uncertainty of $\sigma_\alpha<\alpha/3$ once the angle~$\alpha$ exceeds $\alpha=\ang{3.4}$. We can therefore say that we would be able to detect birefringence angles of $\alpha>\ang{3.4}$ with greater than $3\,\sigma$ significance. However, for an ideal future experiment, the uncertainty is much smaller because the measurements extend to higher multipoles. Hence, we find that for such an ideal experiment, 
we could achieve a $3\,\sigma$ detection for angles of $\alpha>\ang{0.25}$.

This last result demonstrates the best limit that we could achieve in an ideal experiment without using any information from $B$ modes (and hence with no dependence on the angle calibration for the polarimeters). However, this calculation ignores an important systematic uncertainty. A real polarimeter is never \SI{100}{\percent} efficient and hence the polarised signals need to be corrected by some factor that is typically expected to be slightly less than 1. These factors are determined either through laboratory tests of the polarimeters or as part of the calibration of the experiment using data from the sky (see e.g., \cite{leahy2010}). The effect changes the power spectra in different ways. If, for example, one had a polarimeter that measured \SI{99}{\percent} of the incoming polarised signal, then the $TE$ power spectrum would be smaller than expected by a factor of 0.99, while the $EE$ power spectrum would be smaller by the square of this, or 0.98. One sees, therefore, that an uncertainty in the value of this efficiency factor is degenerate with the effects of birefringence in \cref{eq:biref}.

As explained in \cref{sec:intro}, polarisation efficiency and $\cos(2\alpha)$ change the CMB power spectra in the same way. 
For small angles~$\alpha$, $\cos(2\alpha)\approx 1-2\alpha^2$. Hence, we expect that the effect of birefringence will be degenerate with a polarisation efficiency that differs from unity by
an amount $2\alpha^2$. 
Equivalently, the uncertainty on the polarization efficiency corresponds to $2 \sin (2\alpha)\sigma_\alpha$, where $\sigma_\alpha$ is the uncertainty on the birefringence angle. 
Thus, to obtain the most ambitious constraint on the birefringence angle~$\alpha$ coming from $T$ and $E$ we need to determine the polarisation efficiency at a level better than $10^{-4}$.

\section{MCMC}
\label{sec:MCMC}

The \textit{Planck} satellite currently provides the most accurate full-sky temperature anisotropy data and also sensitive measurements of the CMB polarisation using its two sets of instruments, the Low Frequency Instrument (LFI) and the High Frequency Instrument (HFI). Each instrument measured the total intensity and polarisation of photons in several different frequency channels. Various ground-based and in-flight calibration processes were undertaken to characterise the polarimeters and constrain their systematic uncertainties~\cite{leahy2010,rosset2010,planck2013-p02b,planck2013-p03f,planck2014-a06,planck2014-a09,2021break}. 

In this section, we aim to test the method against real \textit{Planck} data. To investigate how well this method constrains cosmic birefringence in real data, we use a Markov chain Monte Carlo (MCMC) approach with the {\textit{Planck}-2018 TT+TE+EE+lowE} likelihood \cite{planck2016-l05} for the standard 6-parameter $\Lambda$CDM cosmology extended by the cosmic birefringence angle~$\alpha$ as a 7th variable parameter.
For this purpose we used the adaptive, speed-hierarchy-aware MCMC sampler (adapted from \textsc{CosmoMC})~\cite{Lewis:2002ah,Lewis:2013hha}, together with \textsc{Cobaya}~\cite{Torrado:2020xyz}, the code interfacing with the \textit{Planck} likelihoods and the \textsc{CAMB} cosmological Boltzmann code~\cite{Lewis:1999bs,Howlett:2012mh}. We modified \textsc{Cobaya} to rotate the CMB power spectra calculated by \textsc{CAMB} according to the birefringence \cref{eq:biref} before passing them on to the likelihood codes.
For post-processing and visualisations we have used the \textsc{GetDist} and \textsc{anesthetic} packages~\cite{getdist,anesthetic}.

As mentioned earlier, the birefringence effect that we discuss here is degenerate with polarisation efficiency. 
Polarisation efficiency is measured for the three most important CMB frequency channels as part of the \textit{Planck}-HFI polarisation calibration. In practice things are a little more complicated than this, since observations from different detectors at different times are co-added to each other to produce temperature and polarisation maps~\cite{planck2016-l05} and the ``polarisation efficiency correction'' parameters for $EE$ at each frequency~$\nu$, $c_\nu^{EE}$ are determined after co-adding data at the power spectrum level and comparing the observed $EE$ power spectra with the theory powers computed from the $\Lambda$CDM best-fit to the $TT$ data. The relevant values at \SIlist{100;143;217}{\giga\hertz} are given by the \textit{Planck} Collaboration~\cite{planck2016-l05}:
\begin{equation}
\begin{aligned}
&c_{100}^{EE}=1.021 \pm 0.010; \\
&c_{143}^{EE}=0.966 \pm 0.010; \\
&c_{217}^{EE}=1.040 \pm 0.010.
\label{eq:poleff}
\end{aligned}
\end{equation}
These polarisation efficiency correction parameters are included in the \textit{Planck} high-$\ell$ likelihood as nuisance parameters. They were fixed to their central values in the 2018 \textit{Planck} Collaboration analyses, since they were deemed to make negligible difference for determining the parameters of the standard 6-parameter $\Lambda$CDM model and some single-parameter extensions~\cite{planck2016-l05}. However, it becomes much more important to include freedom in the values of these parameters when considering the 7-parameter model additionally incorporating the birefringence angle~$\alpha$. In order to account for the degeneracy, we fit the $\Lambda$CDM+$\alpha$ model for three cases: one with flat priors on the polarisation efficiency parameters~$c_\nu^{EE}$ in the interval $(0.8,1.2)$; another with Gaussian priors as given in \cref{eq:poleff}; and one with the polarisation efficiencies fixed to their central values in \cref{eq:poleff}.  

It should be noted that the polarisation efficiency parameters from \cref{eq:poleff} were derived by fitting the data to the $\Lambda$CDM model without birefringence. Therefore, they cannot be used for determining a birefringence angle. One would instead need efficiency parameters that were measured in the laboratory, or determined using the data in some different way (e.g., with foregrounds). Since we do not have such information from \textit{Planck}, we make the assumption that any such constraints would be similar to those in \cref{eq:poleff}.  
Consequently, our analysis in this section does not aim at reporting a constraint on the birefringence angle from \textit{Planck} data, but at highlighting the extent of differences in constraints depending on our prior information on polarisation efficiencies.

\begin{figure}[tb]
    \centering
    \includegraphics[scale= 0.7]{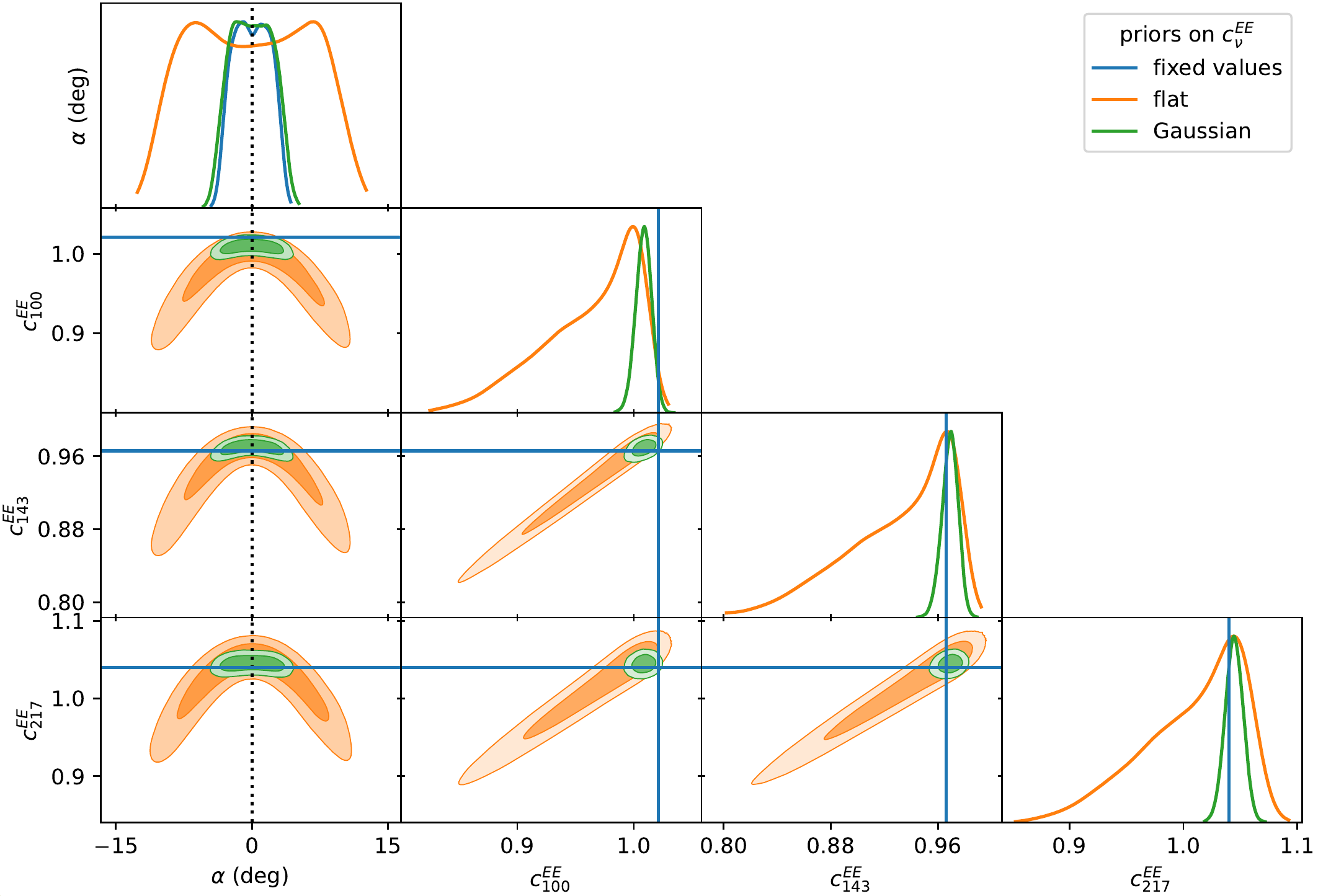}
    \caption{Constraints on the birefringence angle~$\alpha$ and the polarisation efficiency parameters~$c_\nu^{EE}$ at frequencies of $\nu=$~\SIlist{100;143;217}{GHz} for three cases: fixed polarisation efficiency parameters in blue, Gaussian priors on $c_\nu^{EE}$ in green, and flat priors in orange. The black dotted line marks $\alpha=0$.}
    \label{fig:MCMC}
\end{figure}

The resulting posterior distributions of the birefringence angle~$\alpha$ and the polarisation efficiency correction parameters~$c_\nu^{EE}$ are shown in \cref{fig:MCMC}. As can be seen in \cref{eq:biref}, the equations modifying $TT$, $TE$, and $EE$ spectra are symmetric around $\alpha = 0$. Hence, the posterior distributions are also symmetric for positive and negative birefringence angles (as expected, see e.g., \cite{Gruppuso2016}). 

Using a flat prior on the polarisation efficiency correction parameters shows how strongly correlated these parameters are with respect to one another and with the birefringence angel~$\alpha$. Increasing the birefringence angle will reduce the amplitude of our model $EE$ and $TE$ power spectra, as can be seen in \cref{eq:biref}, which then has to be compensated by accordingly lower polarisation efficiency correction parameters. This ultimately ends in a very large uncertainty on all these parameters. For the birefringence angle we obtain a standard deviation of $\sigma_\alpha=\ang{6.2}$ in this case.
Applying tightly-constrained priors like the Gaussian priors in \cref{eq:poleff} or fixing the polarisation efficiency correction parameters breaks the degeneracy and visibly improves constraints on the birefringence angle. In the case of Gaussian priors we obtain a standard deviation of $\sigma_\alpha=\ang{2.1}$ and for fixed polarisation efficiency correction parameters we obtain $\sigma_\alpha=\ang{1.9}$. In contrast to the flat prior, going from Gaussian priors to fixed values results in only a minor improvement on the posterior constraints. 
These results show that having relatively precise measurements of polarisation efficiency significantly improves the constraint on birefringence compared to being uninformed. For future experiments, we need to place tight constraints on these parameters through external calibration steps to be able to check any possible detection of birefringence.

\section{Conclusion}

Constraining cosmic birefringence using CMB power spectra comes with the challenge of distinguishing it from spurious effects caused by systematics. In this paper, we have studied the idea of extracting birefringence from only temperature and $E$\=/mode polarisation data. Although this method gives weaker constraints compared with $B$\=/based limits, we suggest that it can be used as a cross-check for future possible detections based on $B$ modes.

We have used a Fisher matrix analysis to determine the amount of information about cosmic birefringence that we could extract from a \textit{Planck}-like experiment, as well as from an ideal cosmic-variance-limited experiment, using only information from the $TT$, $TE$, and $EE$ power spectra. Our results show that if we could measure $TE$ and $EE$ power spectra to higher multipoles (around $\ell\sim6000$), our constraint on the birefringence angle would be an order of magnitude better than an experiment with \textit{Planck}'s characteristics. 
However, in order to achieve this level of precision we need to calibrate polarisation efficiency to better than the $10^{-4}$ level. This will certainly be challenging, requiring dramatic improvements compared to current capabilities.  However, it may also be possible to separate birefringence from polarisation efficiency effects by using the fact that foregrounds are negligibly affected by cosmological birefringence.

We also have used \textit{Planck} temperature and $E$\=/mode data to fit for the birefringence angle and the polarisation efficiencies simultaneously. We performed MCMC analyses using three different priors on polarisation efficiency parameters: 
(1) an uninformative flat prior, which results in a weak constraint on $\alpha$; 
(2) Gaussian priors with the values determined by the \textit{Planck} Collaboration; 
and (3) fixed values for $c_\nu^{EE}$s, as used by the \textit{Planck} Collaboration in their 2018 analyses. 
Due to the degeneracy, when we used the free polarisation correction parameters, the constraint on the birefringence angle~$\alpha$ degraded substantially. In order to take full advantage of upcoming data and obtain tight constraints on cosmological parameters, it will be necessary to measure and control polarisation efficiency to a challenging level of precision (perhaps using foregrounds). Nonetheless, the $E$\=/mode approach that we have described here represents an additional way of checking any hints of a birefringent signature coming from the more traditional $B$\=/mode methods.

\acknowledgments
We thank Johannes R. Eskilt, Alessandro Gruppuso, and Eiichiro Komatsu for useful comments on an earlier draft. This research was supported by the Natural Sciences and Engineering Research Council of Canada.  
LTH acknowledges funding through a UBC Killam Postdoctoral Research Fellowship. This research was enabled in part by support provided by \textsc{Westgrid} (\url{www.westgrid.ca}) and \textsc{Compute Canada} (\url{www.computecanada.ca}).

\bibliographystyle{JHEP}
\bibliography{refs,Planck_bib,extra}



\end{document}